\documentclass{sigchi}

\usepackage{subcaption}
\usepackage{multirow}
\usepackage[table,xcdraw]{xcolor}

\CopyrightYear{2018}
\setcopyright{acmlicensed}
\conferenceinfo{CHI 2018,}{April 21--26, 2018, Montreal, QC, Canada}
\isbn{978-1-4503-5620-6/18/04}\acmPrice{\$15.00}
\doi{https://doi.org/10.1145/3173574.3174179}

\usepackage{balance}       
\usepackage{graphics}      
\usepackage[T1]{fontenc}   
\usepackage{txfonts}
\usepackage{mathptmx}
\usepackage[pdflang={en-US},pdftex]{hyperref}
\usepackage{color}
\usepackage{booktabs}
\usepackage{textcomp}

\usepackage{microtype}        
\usepackage{ccicons}          

\def\plaintitle{An Experimental Study of Cryptocurrency Market Dynamics}

\def\emptyauthor{}
\def\plainkeywords{Cryptocurrencies; online markets; peer influence; computational social science; online field experiments; digital institutions; design; market design}

\makeatletter
\def\url@leostyle{%
  \@ifundefined{selectfont}{
    \def\UrlFont{\sf}
  }{
    \def\UrlFont{\small\bf\ttfamily}
  }}
\makeatother
\def\pprw{8.5in}
\def\pprh{11in}

\definecolor{linkColor}{RGB}{6,125,233}
\hypersetup{%
  pdftitle={\plaintitle},
  pdfauthor={\emptyauthor},
  pdfkeywords={\plainkeywords},
  pdfdisplaydoctitle=true, 
  bookmarksnumbered,
  pdfstartview={FitH},
  colorlinks,
  citecolor=black,
  filecolor=black,
  linkcolor=black,
  urlcolor=linkColor,
  breaklinks=true,
  hypertexnames=false
}

\begin{document}

\title{\plaintitle}

\numberofauthors{3}
\author{%
  \alignauthor{Peter M. Krafft\thanks{Authors contributed equally. Corresponding author: NDP.}\\
    \affaddr{MIT CSAIL}\\
    \affaddr{Cambridge, MA, USA}\\
    \email{pkrafft@csail.mit.edu}}\\
  \alignauthor{Nicol\'{a}s Della Penna$^*$\\
    \affaddr{Australian National University}\\
    \affaddr{Canberra, Australia}\\
    \email{me@nikete.com}}\\
  \alignauthor{Alex ``Sandy'' Pentland\\
    \affaddr{MIT Media Lab}\\
    \affaddr{Cambridge, MA, USA}\\
    \email{pentland@mit.edu}}\\
}

\maketitle

\begin{abstract}
  As cryptocurrencies gain popularity and credibility, marketplaces
  for cryptocurrencies are growing in importance. Understanding the
  dynamics of these markets can help to assess how viable the
  cryptocurrnency ecosystem is and how design choices affect market
  behavior. One existential threat to cryptocurrencies is dramatic
  fluctuations in traders' willingness to buy or sell.  Using a novel
  experimental methodology, we conducted an online experiment to study
  how susceptible traders in these markets are to peer influence from
  trading behavior.  We created bots that executed over one hundred
  thousand trades costing less than a penny each in 217
  cryptocurrencies over the course of six months.  We find that
  individual ``buy'' actions led to short-term increases in subsequent
  buy-side activity hundreds of times the size of our interventions.
  From a design perspective, we note that the design choices of the
  exchange we study may have promoted this and other peer influence
  effects, which highlights the potential social and economic impact
  of HCI in the design of digital institutions.
\end{abstract}

\category{H.5.m.}{Information Interfaces and Presentation (e.g. HCI)}{Miscellaneous}


\section{Introduction}



Cryptocurrencies, a.k.a. ``cryptocoins'', are rapidly gaining
popularity.  The price and market cap of these assets are touching
all-time highs, with billions of U.S. dollars of value per day
currently being traded in cryptocoins.  Financial institutions are
investing in building digital currency technologies.  Blockchain-based
tech startups are thriving.  As these changes in the cryptocurrency
ecosystem occur, the need to understand the market dynamics of
cryptocoins increases.  At the same time as cryptocurrencies have
gained popularity, their rise has been punctuated by crises.  From the
collapse of Mt. Gox in 2014 to the 2016 hack of Etherium, market
crashes have been a regular occurrence.  Understanding the dynamics of
cryptocurrency markets may allow us to anticipate and avoid future
disruptive events.

One potential threat to cryptocurrencies derives from the speculative
nature of these assets.  Many participants in these markets trade
because they expect one or another cryptocurrency to increase in
value.  Such collective excitement can lead to bubbles and subsequent
market crashes.  The design choices of the online exchanges where
cryptocurrencies are traded may also contribute to these effects if
aspects of available functionality, graphical user interfaces (GUI),
or application programming interfaces (API) promote collective
excitement.  Markets are human artifacts, not natural phenomena, and
therefore a target of design \cite{lampinen2017market}.  In the
present study we strive to better understand the factors that
contribute to collective excitement in cryptocurrencies, and how the
design of cryptocurrency market mechanisms and interfaces may affect
these processes.

Central to these goals is understanding why people at a particular
time decide to invest in a particular technology, product, or idea.
If the asset is new, or information about it has just been released,
investment might be a rational response to the present state of
information \cite{barberis2003survey}.  Other factors could include
authorities endorsing the investment, or big players making noticeably
large bets on it \cite{barber2008all}.  Another hypothesized source of
collective optimism is peer influence among small individual traders
\cite{bikhchandani2000herd,spyrou2013herding,hirshleifer2003herd}.
Understanding these endogenous peer influence effects is especially
important.  If the dynamics of financial markets are heavily affected
by small trades, different solutions may be needed in order to
stabilize the markets.

Peer influence may play a particularly large role in the
cryptocurrency ecosystem due to the highly speculative nature of these
assets.  While in general the intrinsic value of currencies increases
with greater levels of adoption, here we expect that much of the
trading is speculative.  As is characteristic of many new
technologies, there is a great deal of uncertainty around which
cryptocurrencies will eventually be successful, and there has been a
general feeling that some altcoin will become a transformative
financial technology.  Many participants in these marketplaces
therefore are likely hoping to be early investors in ``the next
Bitcoin'', and we can attempt to observe the extent to which the bets
of these traders may be affected by the bets of their peers.


There are several challenges to identifying the effects of small
individual trades in financial markets.  Much of our present knowledge
about financial markets is derived from analysis of observational
data, but observational analyses are subject to confounding
interpretations.  For example, excess correlation in market prices is
often cited as evidence against traders engaging in the purely
rational behavior predicted by the efficient market hypothesis
\cite{lo2002non,malkiel2003efficient}, but these correlations could be
due to delayed reaction or overreaction to news events, as well as
perhaps due to peer influence.  Experimental evidence is desirable
because of these difficulties with observational data.  Laboratory
studies have been conducted in which causal inferences can be made
(e.g.,
\cite{smith1988bubbles,anderson1997information,cipriani2009herd,palan2013review}),
but these studies are limited in their capacity for generalization to
real financial markets.  Since markets are noisy, and the effect of
individual trades is likely to be small, a field experiment in this
area requires a large sample size, which would be expensive to collect
for scientific purposes in traditional financial markets.

We overcome these challenges using a field experiment in an online
marketplace for cryptocurrencies.  Cryptocurrency markets provide a
unique opportunity for field experimentation due to their low
transaction fees; low minimum orders; and free, readily accessible
public APIs.  We created bots to trade in 217 distinct altcoin markets
in an online exchange called Cryptsy.  Each bot monitored a market and
randomly bought or sold the market's associated altcoin at randomly
spaced intervals.  By comparing our buy and sell interventions to
control trials we can estimate the effects that our trades had on
the dynamics of these markets.  In total we conducted hundreds of
thousands tiny trades in these markets over the course of six months,
and this large sample size allows us to test the effects that our
small individual trades have in these live markets.  With an eye
towards design implications, we also conduct an enumerative analysis
of the importance in this context of Cryptsy's design choices.  We
identify the position of Cryptsy in a space of existing and potential
exchange designs, and discuss the possible effects of the dimensions
of its position.  Our analysis reveals that the traders we study are
susceptible to peer influence and highlights how Cryptsy's design
choices might have exacerbated this effect.

\begin{figure}[!ht]
  \centering
    \includegraphics[width=\linewidth]{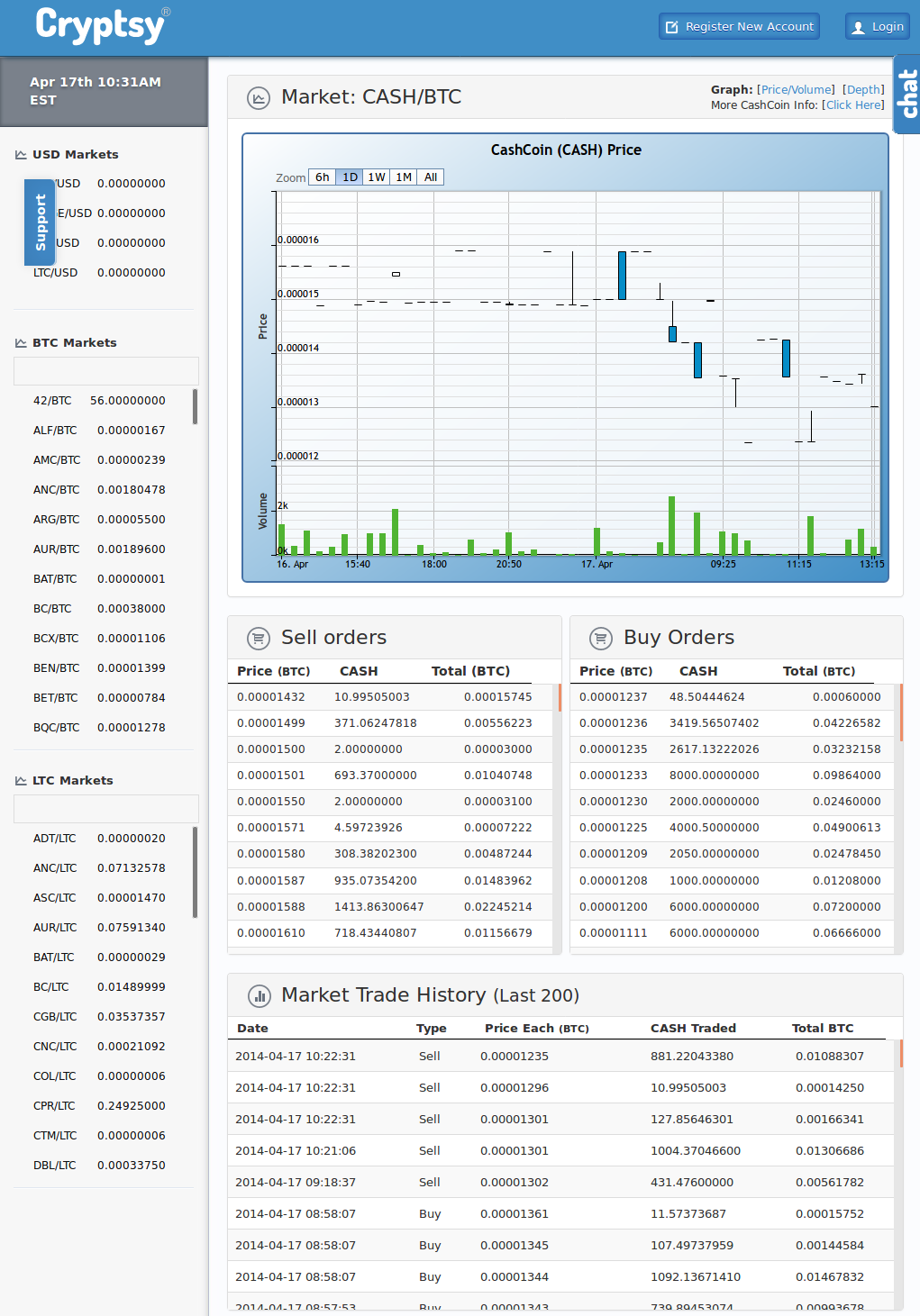}
\caption{A screenshot of the Cryptsy trading platform interface.
  Source: \url{http://archive.is/tDFIw}.}
  \label{fig:screenshot}
\end{figure}

\section{Cryptocurrency Markets}

\subsection{Cryptocurrencies}

Cryptocurrencies are a new type of digital asset that rely on
distributed cryptographic protocols, rather than physical material and
a centralized authority, to operate as currency.  Bitcoin (BTC) was
the first cryptocurrency to gain popularity, but hundreds of
alternative cryptocoins (called ``altcoins'') have since been
introduced.  The current crop of altcoins has been directly inspired
by Bitcoin, and the excitement about Bitcoin frames the hopes and
desires of participants in the marketplace for altcoins.  In
discussions by those who create, promote, and scour such coins, a
desire to not miss out again on being an early investor in the ``next
Bitcoin'' is commonplace.
However, many altcoins represent nothing more than minor changes to
the source code of Bitcoin.  While it might be tempting to dismiss all
such coins as having essentially zero probability of success, some of them do
innovate in non-technical ways.  One case is Auroracoin, which was
a trivial technical modification to Litecoin but was branded as the
official cryptocurrency of Iceland.  Auroracoin at one point had a
market cap of 500 million USD.
Other coins are associated with real technical innovations.  For
example, Ether is used by the Ethereum protocol in order to implement
a fully functional distributed global computer.  Evaluating the
prospective returns from investing in  any of these coins is
difficult and time-consuming, requiring expertise in both cryptography
and economics.

\subsection{Cryptsy}


The platform we use for our experiments, Cryptsy, was a large
cryptocurrency exchange that opened on May 20,
2013
and closed
 in early January
2016.
At the outset of our experiment, Cryptsy claimed over 230,000
registered users.  On the last recorded day of Cryptsy's trading, its daily trading volume was 106,950 USD (248
BTC),
which placed it as the tenth largest cryptocurrency exchanges by
trading volume (of 675 listed) at the
time.
By this time Cryptsy
had 541 trading pairs (including Bitcoin, Litecoin, and fiat markets), which
placed it as the third largest exchange in terms of the total size of
its
marketplace.
Cryptsy was a popular exchange because of the large number of altcoins
it made available for trade, which is also the reason it is uniquely
appropriate for our experiment.  We require a popular exchange with
many coins available.  By the end of 2015, Cryptsy had begun having
well-publicized issues involving users not being able to withdraw
money they had
deposited on the site.
On January 14, 2016 Cryptsy halted all trading.
Our experiment---spanning April 12, 2015 until October 19,
2015---preceded the beginning of the final decline of Cryptsy.


\begin{figure}[!t]
  \centering
  \begin{subfigure}[b]{0.49\linewidth}
    \includegraphics[width=\linewidth]{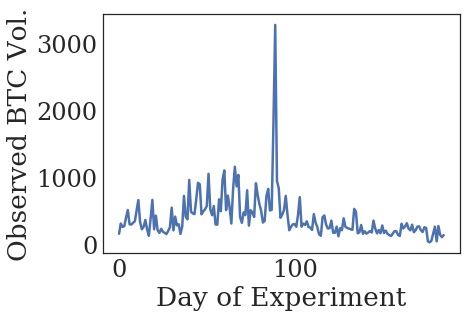}
  \end{subfigure}
  \begin{subfigure}[b]{0.49\linewidth}
    \includegraphics[width=\linewidth]{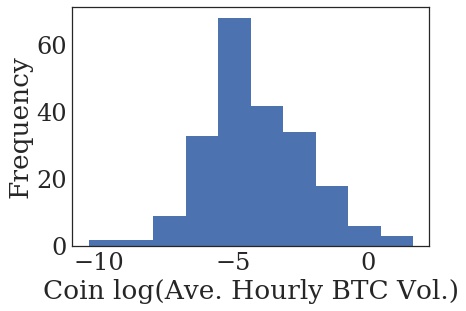}
  \end{subfigure}
  \\
    \begin{subfigure}[b]{0.49\linewidth}
      \includegraphics[width=\linewidth]{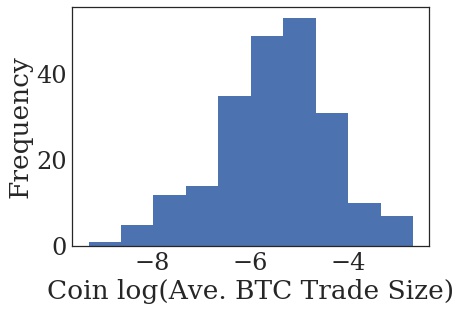}
    \end{subfigure}
    \begin{subfigure}[b]{0.49\linewidth}
      \includegraphics[width=\linewidth]{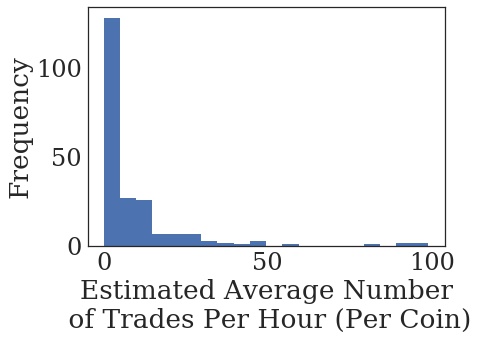}
  \end{subfigure}
\caption{Descriptive statistics of the Cryptsy marketplace.}
\label{fig:descriptives} 
\end{figure}

\subsubsection{Market Mechanism}

Cryptsy implemented a continuous double auction with an open order
book as its trading mechanism.  In a continuous double auction there
is no centralized market maker.  Current asset prices are determined
by the best available prices being offered by any of the traders on
the platform.  A market for a particular cryptocurrency consists of a
set of open ``buy orders'' and a set of open ``sell orders'', all
placed by peers on the site.  A buy order is a request to buy a
quantity of a coin at a price specified in the order.  A sell order is
a request to sell.  The current price to buy, i.e., the ``buy price'',
is given by the lowest priced open sell order.  The current ``sell
price'' is given by the highest priced open buy order.  A transaction
occurs when a new buy order is placed with a price at or above the
lowest sell price, or when a new sell order is placed with a price at
or below the highest buy price.
The minimum denomination on Cryptsy was 1e-8.  At a typical
exchange rate of approximately 200 USD to 1 BTC during the time of our
experiment, 1e-8 BTC corresponded to roughly 0.000002 USD.
The minimum total order size for a single trade varied across
coins and over time, but was typically 1e-7 BTC, and the transaction
fees were negligible.

\subsubsection{Interfaces}

Cryptsy had both a graphical user interface (GUI) and an application
programming interface (API).  The GUI, pictured in Figure
\ref{fig:screenshot}, allowed users to browse through USD, BTC, and
Litecoin (LTC) markets.  The site allowed users to view a standalone
list of all the markets, which could be sorted by volume or price
(recent price, 24-hour high price, or 24-hour low price), and the site
allowed users to view details of the specific coin markets.  All
current coin prices for each of those markets were displayed in panels
on the left-hand side of the screen in alphabetical order by the coin
name.  For the BTC and LTC markets, these panels also displayed green
or red marks when the prices of each coin had recently moved up or
down.  These marks accumulated during an
idle web session.  The single most recent BTC and LTC market price
changes were displayed at the tops of these panels.

In addition to showing all of the open buy and sell orders, Cryptsy
also made prior transactions visible to users.  Prior transactions
were shown both in a list and as a chart.  The list included the sizes,
prices, and times of the last 200 transactions, reverse ordered by
time.  The chart displayed an interactive summary of prior
transactions.  Each ``tick'' on the chart visualized the highest
transacted price, the lowest transacted price, the open price, and the
close price over the duration of a certain time interval.  The total
traded volume over the time duration of each tick was also shown.  The
time interval used for display depended on the time granularity at
which the chart was being viewed.  The chart could display the price
trends over the last 6 hours, day, week, or month, or over all time.




\subsubsection{Descriptive Statistics}



Figure \ref{fig:descriptives} summarizes various descriptive
statistics of the Cryptsy marketplace.  We observe a mean daily
trading volume of approximately 400 BTC per day on Cryptsy.  This
quantity places the mean daily trading volume on Cryptsy to be at
least in the tens of thousands of USD per day during our experiment.
The mean daily trading volume remained relatively constant over the
course of our experiment, with some periods of higher volume.  There
was substantial heterogeneity in the volume of each coin.  Most coins
have only on the order of 10 or 100 USD being traded in their markets
per day, while a few have tens of thousands.  The average size of
observed trades also varies widely across coins.  Across all coins,
average trade sizes tend to be in the range of tens of cents to a few
dollars.  The maximum trade size we use in our experiments of 5e-6 BTC
is in the bottom 8\% of the distribution of observed trade sizes,
while our minimum trade size of 5e-7 is in the bottom 1\%.  Using the
average BTC size of trades and the average hourly BTC volume per coin,
we can also estimate the average number of trades per hour per coin.
These estimates indicate that most coins tend have a handful of trades
per hour at the times we execute our interventions.

\begin{table*}[!t]
  \centering
  \footnotesize
  \begin{tabular}{|r|r|r|r|r|r|r|r|r|}
    \hline
    \textbf{Time} & \textbf{Condition} & \textbf{Dependent Var.} & \textbf{$n$ Control} & \textbf{$n$ Treat} & \textbf{Control Mean} & \textbf{Mean Effect} & \textbf{$t$-stat} & \textbf{$p$-value}\\\hline
    15 Min. & Buy & Buy Prob. & 25483 & 25602 & 0.279 & 0.019 & 4.79 & 2.96e-05\\\hline
    15 Min. & Buy & \% Buy Vol. & 24321 & 24313 & 0.290 & 0.017 & 4.40 & 1.97e-04\\\hline
    15 Min. & Buy & Trade Prob. & 52050 & 51314 & 0.490 & 0.009 & 3.00 & 4.81e-02\\\hline
    15 Min. & Sell & Sell Prob. & 25483 & 25987 & 0.721 & -0.006 & -1.44 & 1.00e+00\\\hline
    15 Min. & Sell & \% Sell Vol. & 24321 & 24660 & 0.710 & -0.005 & -1.31 & 1.00e+00\\\hline
    15 Min. & Sell & Trade Prob. & 52050 & 51727 & 0.490 & 0.013 & 4.12 & 6.71e-04\\\hline
    30 Min. & Buy & Buy Prob. & 23647 & 23871 & 0.278 & 0.003 & 0.83 & 1.00e+00\\\hline
    30 Min. & Buy & \% Buy Vol. & 23583 & 23802 & 0.291 & 0.005 & 1.33 & 1.00e+00\\\hline
    30 Min. & Buy & Trade Prob. & 52049 & 51312 & 0.454 & 0.011 & 3.51 & 7.98e-03\\\hline
    30 Min. & Sell & Sell Prob. & 23647 & 23809 & 0.722 & 0.001 & 0.15 & 1.00e+00\\\hline
    30 Min. & Sell & \% Sell Vol. & 23583 & 23735 & 0.709 & 0.002 & 0.58 & 1.00e+00\\\hline
    30 Min. & Sell & Trade Prob. & 52049 & 51724 & 0.454 & 0.006 & 1.94 & 9.53e-01\\\hline
    60 Min. & Buy & Buy Prob. & 31065 & 31118 & 0.274 & 0.003 & 0.76 & 1.00e+00\\\hline
    60 Min. & Buy & \% Buy Vol. & 30984 & 31044 & 0.289 & 0.001 & 0.33 & 1.00e+00\\\hline
    60 Min. & Buy & Trade Prob. & 52030 & 51288 & 0.597 & 0.010 & 3.18 & 2.70e-02\\\hline
    60 Min. & Sell & Sell Prob. & 31065 & 31351 & 0.726 & 0.000 & 0.14 & 1.00e+00\\\hline
    60 Min. & Sell & \% Sell Vol. & 30984 & 31275 & 0.711 & -0.003 & -1.00 & 1.00e+00\\\hline
    60 Min. & Sell & Trade Prob. & 52030 & 51713 & 0.597 & 0.009 & 3.02 & 4.50e-02\\\hline
    \end{tabular}
  \caption{t-tests for our three main dependent variables at each
    monitor time associated with our interventions.  All statistics
    were computed on our confirmatory dataset.  $p$-values are
    two-sided and Bonferroni corrected for 18 tests.  We have
    differing numbers of observations between our probability and
    percentage statistics for like conditions at identical monitor
    events due to differences in sensitivity to trade execution
    latency in how we implemented these statistics.}
\label{table:ttests}
\end{table*}

\begin{table}[!t]
  \centering
  \footnotesize
  \begin{tabular}{|r|r|r|r|r|}

    \hline

    \textbf{Dep. Var.} & \textbf{Independent Var.} & \textbf{Coef.} & \textbf{$t$-stat} & \textbf{$p$-value}\\

    \hline

    Buy Prob. & Buy Treat. & 0.016 & 4.47 & 1.4e-04 \\\hline

    Buy Prob. & Buy Treat.*Time 2 & -0.016 & -3.02 & 4.48e-02 \\\hline

    Buy Prob. & Buy Treat.*Time 3 & -0.015 & -3.15 & 2.92e-02 \\\hline

    Buy Prob. & Sell Treat. & 0.004 & 1.01 & 1.00e+00 \\\hline

    Buy Prob. & Sell Treat.*Time 2 & -0.004 & -0.72 & 1.00e+00 \\\hline

    Buy Prob. & Sell Treat.*Time 3 & -0.003 & -0.61 & 1.00e+00 \\\hline

    \% Buy Vol. & Buy Treat. & 0.014 & 4.14 & 6.39e-04 \\\hline

    \% Buy Vol. & Buy Treat.*Time 2 & -0.012 & -2.46 & 2.53e-01 \\\hline

    \% Buy Vol. & Buy Treat.*Time 3 & -0.015 & -3.30 & 1.75e-02 \\\hline

    \% Buy Vol. & Sell Treat. & 0.003 & 0.98 & 1.00e+00 \\\hline

    \% Buy Vol. & Sell Treat.*Time 2 & -0.005 & -1.03 & 1.00e+00 \\\hline

    \% Buy Vol. & Sell Treat.*Time 3 & 0.001 & 0.21 & 1.00e+00 \\\hline

    Trade Prob. & Buy Treat. & 0.006 & 2.28 & 4.05e-01 \\\hline

    Trade Prob. & Buy Treat.*Time 2 & 0.002 & 0.41 & 1.00e+00 \\\hline

    Trade Prob. & Buy Treat.*Time 3 & 0.000 & 0.08 & 1.00e+00 \\\hline

    Trade Prob. & Sell Treat. & 0.011 & 4.11 & 7.1e-04 \\\hline

    Trade Prob. & Sell Treat.*Time 2 & -0.007 & -1.79 & 1.00e+00 \\\hline

    Trade Prob. & Sell Treat.*Time 3 & -0.004 & -0.98 & 1.00e+00 \\\hline

    \end{tabular}
\caption{Linear regressions for our three main dependent variables,
  including fixed effects for each coin, control variables for market
  state, and cluster-robust standard errors.  All statistics were
  computed on our confirmatory set.  $p$-values are two-sided
  and Bonferroni corrected for 18 tests.  The significant negative
  effects at Time 2 and Time 3 indicate that the positive peer
  influence effect of our buy interventions diminishes over time.}
\label{table:regressions}
\end{table}

\section{Experiment}


\subsection{Procedure}

We conducted 310,222 randomly spaced trials in the Cryptsy exchange
over the course of six months.  Before our experiment, we purchased
0.002 BTC worth each of 217 different altcoins available on Cryptsy.
We then created bots that monitored and periodically traded in each of
these altcoins' markets in parallel.  Each bot waited a random amount
of time, between one and two hours, at the beginning of the experiment
before conducting a first trade.  Then, assuming there had been at
least one trade in the last hour on a coin, the bot for that coin
recorded the current market state and randomly chose to either buy a
random small amount of the coin, sell the coin, or do nothing (as a
control condition).  The trade sizes were chosen uniformly at random
between 5e-7 and 5e-6 BTC.  Each bot then observed its coin's market
state at 15 minutes, 30 minutes, and 60 minutes following the trial,
after which the bot waited a random amount of time between 0 and 60
minutes before engaging in another trial and repeating.  As a part of
monitoring the market state, each bot recorded the details of the most
recent trade in the market, as well as the total buy-side and
sell-side volume since the bot's intervention.  We conducted a power
analysis based on a pilot study to determine the length of time to run
our experiment.

\subsection{Dependent Variables}

We examine three dependent variables in our main analyses.  In order
to measure peer influence on trade direction, we use two statistics.
The first statistic is the probability that the last trade observed on
a coin is a buy as opposed to a sell, conditional on there being a
trade in the observation period of the statistic.  This statistic uses
the transaction type of the last transaction that occurred in the time
period associated with a given monitor event. The statistic aggregates
binary indicator variables that check whether the last observed
transaction types were buys or sells.  For this statistic, we omit
monitor events where we have not observed a trade after our own
initial trade or after the last monitor event.  The second statistic
is the proportion of trading volume on the buy-side or sell-side after
our interventions.  To compute this statistic, we calculate the total
volume of trades of a given coin that occurred in the time period
associated with a given monitor event, and we calculate the fraction
of that volume associated with buy or sell trades.  Once again, this
statistic is undefined when there has been no new trading activity at
a given monitor event.  Our statistical tests compare the values of
these statistics across the treatment and control trials in each
monitor period.  Peer influence in buying or selling would be
indicated by buy treatments leading to higher buy-side activity as
measured by these statistics, or by sell treatments leading to higher
sell-side activity.  A third statistic, the probability of observing
any trade at all in each monitor period, allows us to test for peer
influence in overall trading activity in addition to trade direction.
We compute each of these statistics in three mutually exclusive
periods: in the 15 minutes after our interventions, the following 15
minutes, and the following 30 minutes.




\begin{figure}[t]
  \centering
  \begin{subfigure}[b]{\linewidth}
    \includegraphics[width=0.49\linewidth]{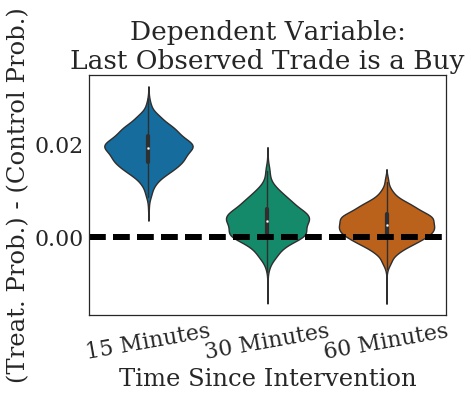}
    \includegraphics[width=0.49\linewidth]{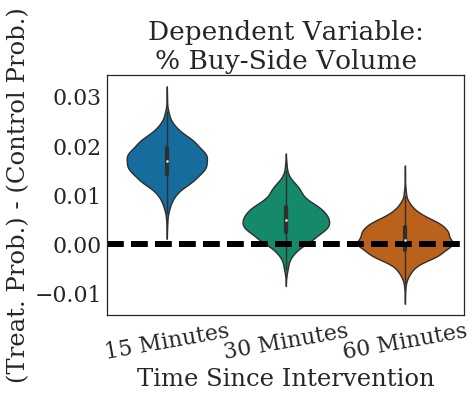}
    \caption{Buy-side Interventions}    
  \end{subfigure}
  \centering
  \begin{subfigure}[b]{\linewidth}
    \includegraphics[width=0.49\linewidth]{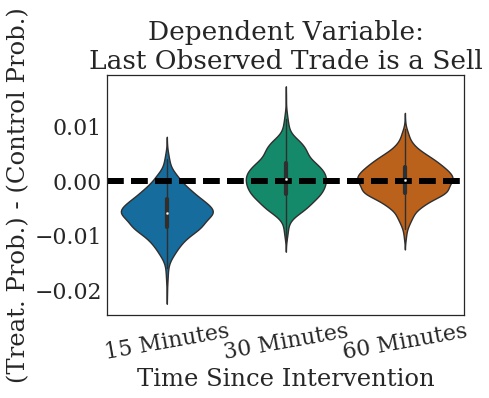}
    \includegraphics[width=0.49\linewidth]{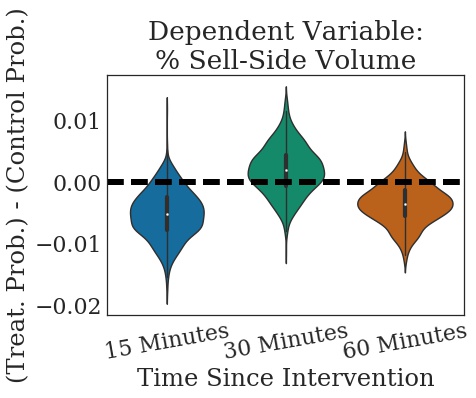}
        \caption{Sell-side Interventions}
  \end{subfigure}
  \caption{Bootstrap test statistic distributions for two dependent
    variables as a function of time after our
    interventions.}\label{fig:results}
  \end{figure}

\section{Results}

\subsection{Statistical Analysis}

\begin{figure}[t]
  \centering
  \begin{subfigure}[b]{0.45\linewidth}
    \includegraphics[width=\linewidth]{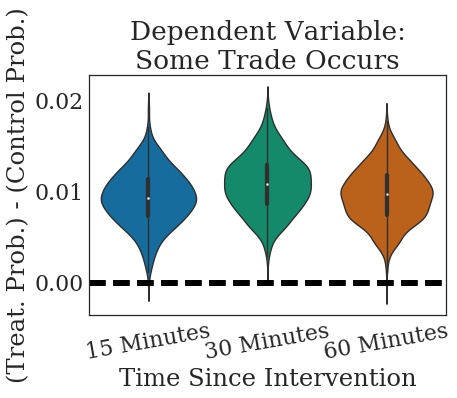}
    \caption{Buy-side Interventions}    
  \end{subfigure}
  \centering
  \begin{subfigure}[b]{0.45\linewidth}
    \includegraphics[width=\linewidth]{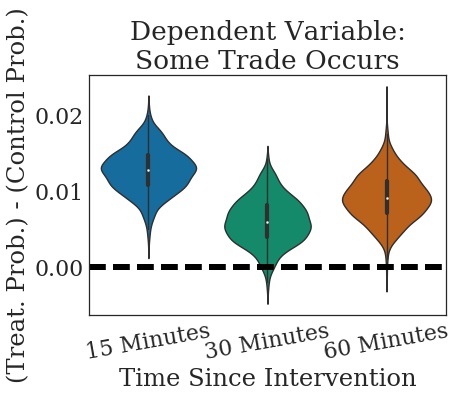}
        \caption{Sell-side Interventions}
  \end{subfigure}
  \caption{Bootstrap test statistic distributions for our final
    dependent variable as a function of time after our
    interventions.}\label{fig:null-results}
  \end{figure}

For a basic analysis, we use t-tests to compare the average values of
our dependent variables in our buy and sell trials to the average
values in our control trials.  The results of these tests are given in
Table \ref{table:ttests} and visualized in Figures \ref{fig:results}
and \ref{fig:null-results}.  We find significant effects of our buy
interventions on buy probability and percent buy-side volume at the
15-minute monitor event ($p < 0.001$), and significant effects of buy
and sell interventions on overall trading activity at all monitor
events ($p < 0.05$) except the 30-minute monitor for sell
interventions.

In addition to these t-tests, we perform
a more robust analysis that combines the analysis of our buy and sell
treatments across monitor times, while also controlling for potential
dependence between observations on identical coins and potential
dependence due to correlations in time.  We perform linear
fixed-effects regressions that include intercepts for each coin, and
include regressor variables summarizing the market state to control
for temporal dependence in the form of spillover effects.  The
variables we use to control for a coin's market state are: six
indicator variables for whether the best buy or sell order in the
coin's order book are above, below, or at the coin's last transaction
price---these variables indicate how a new trade would affect the
price trend of that coin; an indicator variable for whether the
last trade before our intervention was a buy or a sell; a
continuous variable giving the percentage of buy-side versus sell-side
volume in the hour before the intervention; and a continuous
variable giving the logarithm of the coin market volume in the hour
before our intervention divided by the average hourly volume on that
coin.  These regressions also include two indicator variables for
whether an observation is associated with a buy trial or a sell trial
(versus a control trial, which is absorbed into the intercept terms),
and the regressions include interaction terms between these trial
indicators and indicator variables for each observation's associated
monitor number.  We use cluster-robust White standard errors
\cite{white1980heteroskedasticity}, which adjust the model standard
errors to account for heteroskedasticity and correlation in error
terms within each altcoin's observations.  The results of these
regression are given in Table \ref{table:regressions}.  These results
confirm the results of our t-tests.  The effect of our buy treatments
on buy trade probability and buy-side volume percentage remain
significant at the 15-minute monitor event ($p < 0.001$).  The effect
of our buy interventions on the probability of observing a trade at
all is marginally significant, being significant at the 0.05 level
before we correct for multiple comparison but not afterwards.  The
effect of our sell interventions on the probability of observing a
trade at all remains significant ($p < 0.001$).  In the appendix we
examine the randomization validity of our experiment, the extent of
selection bias we might have suffered, and heterogeneity in the
effects of our interventions across coins.

\subsection{Peer Influence}


Our results indicate strong evidence for peer influence on our buy
interventions in these markets.  The probability the last observed
trade is a buy 15 minutes after a buy intervention is 30\%---an
increase of 2\% above the control probability of 28\%.  The average
percentage buy-side BTC volume is 2\% higher than the 29\% we observe
in control trials during these 15 minute windows.  Our interventions
also led to overall increases in trading activity.  The probability of
observing a trade at all within 15 minutes after our interventions
increased from 49\% to approximately 50\% after buy or sell
interventions.

The aggregate effects of slightly higher percentages of buy-side
activity and slightly higher overall levels of trading, in combination
with a fat-tailed distribution of trading volume, accumulate into a
large 7\% average increase in total buy-side BTC volume after our buy
interventions as compared to control trials (MWU-test $p = 0.0011$; we
use an MWU-test since the distributions are fat-tailed, $MAE/MSE =
0.273$ as compared to $0.799$ of the normal distribution).  In total,
we observe approximately 16,000 USD additional buy-side trading in our
buy intervention trials as compared to the control trials.  Even
though we conducted slightly fewer buy interventions than there were
control trials, the sum of all buy-side volume immediately following
our buy interventions was 1513 BTC, as compared to 1430 BTC after the
control trials (and 1399 after sell trials).  This difference of 83
BTC is large compared to the total size of all of our interventions,
which was approximately 0.14 BTC.  The total effects of our buy
interventions were approximately 500 times larger than their cost.




\subsection{Asymmetric Null Effect}

We do not observe a symmetric peer influence effect from sell
interventions.  Sell interventions had no detectable effect on the
proportion of future sell-side trading.  There are multiple potential
causes of this asymmetry.  Since we executed roughly the same number
of buy and sell interventions, and since the dependent variables in
each case have the same variance, the asymmetry cannot be due to a
difference in statistical power.  Therefore the difference must be due
to differences in how the market behaves following buy versus sell
events.  Notably, asymmetric peer influence effects of this sort have
been observed in online social recommendation systems in which upvotes
lead to peer influence but not downvotes \cite{muchnik2013social}.
However, other contextual factors could be causing the difference in
our case.  One possibility could be the fact that sell actions do not
decrease prices in these markets as frequently as buy actions raise
prices (sell actions only decrease prices 38\% of the time, compared
to the 78\% of buy actions), and hence sell treatments may not lead to
the same momentum effects.  Another potential factor is that
conducting a sell-side trade requires having holdings in the asset
since short-selling was not implemented on Cryptsy.  Therefore the
sell effects could be weaker due to a smaller population of peers with
the capability to sell.


\subsection{Temporal Trends}

The peer influence effects we observe after buy interventions do not
lead to detectable permanent shifts in market dynamics.  By 30 minutes
after our interventions there is no longer a detectable peer influence
effect on trade direction.  This diminishing effect over time could be
due to the fact that natural variation in subsequent trading after our
interventions consists of a mix of buying and selling, which is
variance that could dampen the effects of our buy trades.  The overall
excitation effects we observe are persistent over time.  The
probability of observing a trade between a half hour and an hour after
either a buy or a sell intervention remains over 1\% higher than the
baseline probability in the control condition, which in this time
period is 60\%.

\begin{figure}[t]
  \centering
    \includegraphics[width=0.49\linewidth]{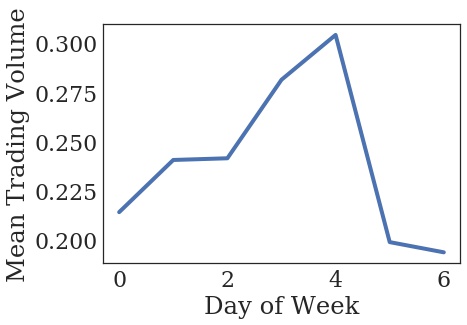}
    \includegraphics[width=0.49\linewidth]{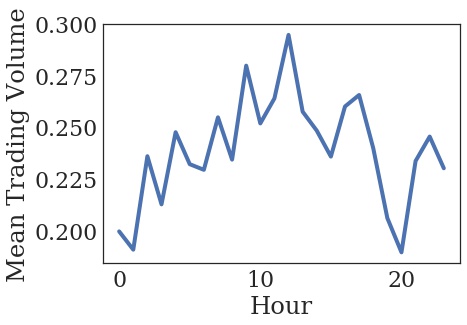}
  \caption{Plots of average hourly and daily volume.}
  \label{fig:time-vol}
\end{figure}


\subsection{Market Composition}

The interpretation of our results depends strongly on what sorts of
participants compose the Cryptsy marketplace.  If there are many
active bots, or if our interventions disproportionately affect bots,
then the implications of our experiment for the potential effects of
exchange design choices would differ.  In order to check if our
results are due mainly to human activity, we conduct a regression
analysis including an independent variable that indicates whether each
intervention was executed during the hours of the NY Stock Exchange
(9am to 4pm Eastern, Monday-Friday) and an interaction term between
this variable and the experimental condition.  Here we compare buy
versus sell interventions directly, rather than each to the control
condition, to maximize our statistical power.  We do observe cyclical
trends overall in average volume during U.S. work hours, as shown in
Figure \ref{fig:time-vol}, which indicates that the markets likely
have substantial human trading.  However, our analysis is inconclusive
since we detect no significant interaction in our regression.  In
exploratory analysis, we obtained similarly inconclusive results using
a variety of independent variables, including holidays, weekdays, and
hour of the day.  We therefore take our null result in this analysis
to be weak evidence that bots are contributing meaningfully to our
results.

\section{Discussion}



\begin{table*}[!htp]
  \footnotesize
  \centering
  \begin{tabular}{|l|l|l|l|l|l|l|}
    \hline
    & \multicolumn{3}{c|}{GUI}                                                                                                     & \multicolumn{3}{c|}{API}                                                       \\ \hline
    & Bitfinex                                        & Bitstamp                                        & Cryptsy                  & Bitfinex                 & Bitstamp                 & Cryptsy                  \\ \hline
    Ticker chart                       & \cellcolor[HTML]{56B4E9}{\color[HTML]{56B4E9} } & \cellcolor[HTML]{56B4E9}                        & \cellcolor[HTML]{56B4E9} &                          &                          &                          \\ \hline
    Prominent ticker chart             & \cellcolor[HTML]{56B4E9}                        & \cellcolor[HTML]{56B4E9}                        & \cellcolor[HTML]{56B4E9} &                          &                          &                          \\ \hline
    Interactive ticker chart           & \cellcolor[HTML]{56B4E9}                        & \cellcolor[HTML]{56B4E9}                        & \cellcolor[HTML]{E69F00} &                          &                          &                          \\ \hline
    Order book visualization           & \cellcolor[HTML]{56B4E9}                        & \cellcolor[HTML]{56B4E9}                        & \cellcolor[HTML]{E69F00} &                          &                          &                          \\ \hline
    Customizable chart colors          & \cellcolor[HTML]{56B4E9}                        & \cellcolor[HTML]{E69F00}                        & \cellcolor[HTML]{E69F00} &                          &                          &                          \\ \hline
    Audio representations of activity  & \cellcolor[HTML]{E69F00}                        & \cellcolor[HTML]{56B4E9}                        & \cellcolor[HTML]{E69F00} &                          &                          &                          \\ \hline
    Quick buy/sell buttons             & \cellcolor[HTML]{E69F00}                        & \cellcolor[HTML]{56B4E9}                        & \cellcolor[HTML]{E69F00} &                          &                          &                          \\ \hline
    Page with coin statistics compared & \cellcolor[HTML]{E69F00}                        & \cellcolor[HTML]{E69F00}                        & \cellcolor[HTML]{56B4E9} &                          &                          &                          \\ \hline
    Active markets/price changes highlighted         & \cellcolor[HTML]{E69F00}{\color[HTML]{E69F00} } & \cellcolor[HTML]{E69F00}{\color[HTML]{E69F00} } & \cellcolor[HTML]{56B4E9} &                          &                          &                          \\ \hline
    Moderated forums                   & \cellcolor[HTML]{E69F00}                        & \cellcolor[HTML]{E69F00}                        & \cellcolor[HTML]{E69F00} &                          &                          &                          \\ \hline
    Coin information pages             & \cellcolor[HTML]{E69F00}                        & \cellcolor[HTML]{E69F00}                        & \cellcolor[HTML]{E69F00} &                          &                          &                          \\ \hline
    Your trades and orders             & \cellcolor[HTML]{56B4E9}                        & \cellcolor[HTML]{56B4E9}                        & \cellcolor[HTML]{56B4E9} & \cellcolor[HTML]{56B4E9} & \cellcolor[HTML]{56B4E9} & \cellcolor[HTML]{56B4E9} \\ \hline
    Recent trades displayed            & \cellcolor[HTML]{56B4E9}                        & \cellcolor[HTML]{56B4E9}                        & \cellcolor[HTML]{56B4E9} & \cellcolor[HTML]{56B4E9} & \cellcolor[HTML]{56B4E9} & \cellcolor[HTML]{56B4E9} \\ \hline
    Live updating                      & \cellcolor[HTML]{56B4E9}                        & \cellcolor[HTML]{56B4E9}                        & \cellcolor[HTML]{E69F00} & \cellcolor[HTML]{56B4E9} & \cellcolor[HTML]{56B4E9} & \cellcolor[HTML]{56B4E9} \\ \hline
    Basic market summary statistics displayed  & \cellcolor[HTML]{56B4E9}                        & \cellcolor[HTML]{56B4E9}                        & \cellcolor[HTML]{56B4E9} & \cellcolor[HTML]{56B4E9} & \cellcolor[HTML]{56B4E9} & \cellcolor[HTML]{56B4E9} \\ \hline
    Summaries of all markets displayed on page      & \cellcolor[HTML]{56B4E9}                        & \cellcolor[HTML]{E69F00}                        & \cellcolor[HTML]{56B4E9} & \cellcolor[HTML]{E69F00} & \cellcolor[HTML]{E69F00} & \cellcolor[HTML]{56B4E9} \\ \hline
    Market indicators available                  & \cellcolor[HTML]{56B4E9}                        & \cellcolor[HTML]{56B4E9}                        & \cellcolor[HTML]{E69F00} & \cellcolor[HTML]{E69F00} & \cellcolor[HTML]{E69F00} & \cellcolor[HTML]{E69F00} \\ \hline
    Communication channel              & \cellcolor[HTML]{56B4E9}                        & \cellcolor[HTML]{E69F00}                        & \cellcolor[HTML]{56B4E9} & \cellcolor[HTML]{E69F00} & \cellcolor[HTML]{E69F00} & \cellcolor[HTML]{E69F00} \\ \hline
    Trader identities available        & \cellcolor[HTML]{E69F00}                        & \cellcolor[HTML]{E69F00}                        & \cellcolor[HTML]{E69F00} & \cellcolor[HTML]{E69F00} & \cellcolor[HTML]{E69F00} & \cellcolor[HTML]{E69F00} \\ \hline
    ``Recommended coins''                & \cellcolor[HTML]{E69F00}                        & \cellcolor[HTML]{E69F00}                        & \cellcolor[HTML]{E69F00} & \cellcolor[HTML]{E69F00} & \cellcolor[HTML]{E69F00} & \cellcolor[HTML]{E69F00} \\ \hline
    Published rate limits              &                                                 &                                                 &                          & \cellcolor[HTML]{56B4E9} & \cellcolor[HTML]{56B4E9} & \cellcolor[HTML]{E69F00} \\ \hline
    \end{tabular}
    \caption{Comparison of the graphical user interfaces (GUIs) and
    application programming interfaces (APIs) of three cryptocurrency
    exchanges.  Blue indicates the feature is present.  Orange
    indicates the feature is absent.  White indicates the feature is
    not applicable.}
    \label{table:interfaces}
  \end{table*}

\begin{table}[!htp]
  \footnotesize
  \centering
  \begin{tabular}{|l|c|c|c|}
    \hline
    & Bitfinex                                        & Bitstamp                 & Cryptsy                  \\ \hline
    Coins                     & 25                                              & 5                        & 217                      \\ \hline
    Continuous double auction & \cellcolor[HTML]{56B4E9}                        & \cellcolor[HTML]{56B4E9} & \cellcolor[HTML]{56B4E9} \\ \hline
    Open order book           & \cellcolor[HTML]{56B4E9}                        & \cellcolor[HTML]{56B4E9} & \cellcolor[HTML]{56B4E9} \\ \hline
    Open trade history        & \cellcolor[HTML]{56B4E9}                        & \cellcolor[HTML]{56B4E9} & \cellcolor[HTML]{56B4E9} \\ \hline
    Spot trading              & \cellcolor[HTML]{56B4E9}{\color[HTML]{56B4E9} } & \cellcolor[HTML]{56B4E9} & \cellcolor[HTML]{56B4E9} \\ \hline
    Margin trading            & \cellcolor[HTML]{56B4E9}                        & \cellcolor[HTML]{E69F00} & \cellcolor[HTML]{E69F00} \\ \hline
    Short-selling             & \cellcolor[HTML]{56B4E9}                        & \cellcolor[HTML]{E69F00} & \cellcolor[HTML]{E69F00} \\ \hline
    Market-making             & \cellcolor[HTML]{E69F00}                        & \cellcolor[HTML]{E69F00} & \cellcolor[HTML]{E69F00} \\ \hline
  \end{tabular}
  \caption{Comparison of cryptocurrency exchange market mechanisms.
    The color code is the same as in Table \ref{table:interfaces}.}
      \label{table:markets}
\end{table}

\subsection{Mechanisms}

We now discuss potential mechanisms underlying the results of our
experiment.  We have limited ability to distinguish between potential
causes due to the scope our data and experiment.  In lieu of being
able to provide evidence to favor one hypothesis over another, we
attempt to enumerate possible explanations.

\subsubsection{Potential Ultimate Causes}

In the analysis of our experiment we attempted to identify to what
extent our results might have been attributable to bot trading.  We
have a limited ability to do so since we cannot observe the identities
of traders or which trades are executed via the GUI versus via the
API, and we know of no major changes to the trading page interface
during our experiment that could afford quasi-experimental analysis.
In any case, our results are ultimately due to human behavior,
whether:
\begin{itemize}
\item Direct human decision-making in GUI trading.
  \item Human planning, as manifest in what bots are programmed.
\end{itemize}
These two possibilities each implicate different mediating mechanisms
and moderating contextual factors.

\subsubsection{Potential Behavioral Mechanisms}

There are a few plausible simple behavioral mechanisms that could
underlie our results:
\begin{itemize}
\item Explicitly copying buy trades.
\item Buying on an increasing price trend, i.e. momentum.
\item Buying salient coins, such as coins with recent activity.
\end{itemize}
Since trade history is openly available on Cryptsy, a simple
explanation is that some fraction of traders decide to buy if there
has recently been a buy from another trader.  Another possibility is
that traders are paying attention to price increases, rather than
explicitly copying buy actions.  Buy actions raised the last
transacted price 78\% of the time during our experiment.
Another possibility is that people buy salient coins. The frequent
price increases from buy actions were reflected in the charts that
display summaries of the prices of each coin.  Peer influence from buy
actions may therefore be due to buying these salient coins.
The first two mechanisms could easily be implemented either by people
directly or through bots.  The last could be implemented by bots, but
would more plausibly be implemented by people since the API does not
highlight price changes.  More sophisticated mechanisms are also a
possibility if traders conduct more complex calculations that result
in individual buy trades having an overall marginal positive peer
influence effect.

\subsubsection{Potential Moderating Contextual Factors}

There are also several plausible contextual factors that might
moderate the effects we observe:
\begin{itemize}
\item Interface and market features
\item Uncertainty of cryptocurrencies
\item Aspects of the community disposition
\item Size of Cryptsy
\end{itemize}
We discuss interface and market features in more detail below.  Other
contextual factors that might affect the generalizability of our
results are the highly speculative nature of cryptocurrencies,
peculiarities of the composition of traders on Cryptsy, or the size of
the Cryptsy marketplace.  We expect that speculation, amateur trading,
and a relatively small marketplace in which our interventions are more
visible might have promoted peer influence in our study.

\subsection{Cryptocurrency Exchange Comparison}



The website CryptoCoinCharts currently lists 125 online exchanges where bitcoins
and various altcoins can be traded.  In this section we discuss common
features of these exchanges as well as other features that could
plausibly be implemented in order to explore a design space of
cryptocurrency exchanges and to provide context for our experiment.
This analysis illustrates that the structure of Cryptsy is relatively
representative of the structure of currently popular cryptocurrency
exchanges, and provides a point of reference for a discussion of
plausible interface effects in our experiment.

We compare Cryptsy to the two highest-volume cryptocurrecy exchanges
at the time of writing: Bitfinex and Bitstamp.  Tables
\ref{table:interfaces} and \ref{table:markets} summarize key features
of these exchanges.  The three exchanges are similar in many ways, but
also have some important differences.  Common to all the exchanges is
the use of a continuous double auction market mechanism, open order
books and trading history, spot trading functionality, and a prominent
ticker chart visualizing recent price changes.  One notable difference
in market mechanisms is that Bitfinex implements margin trading, which
allows for leveraged trades and short-selling.  Notable differences in
the interfaces are that Cryptsy did not include charts for market
indicators; Bitfinex and Bitstamp do not highlight outside market
activity on an individual coin's trading page; and Bitstamp does not
have a communication channel for traders---Cryptsy had its own chat
room, while Bitfinex links directly to a separate site, TradingView,
from its platform.  None of the exchanges implemented market makers,
and none of the exchanges implemented common features of other types
of online platforms, such as moderated forums, information pages,
deanonyimized user activity, or recommendations.

\subsection{Interface Effects}

Given the design space we outlined in our comparison of cryptocurrency
exchanges, we can hypothesize about which interface and market
features might enable or promote peer influence.  The design choices
of the GUI clearly could impact human trading, and the affordances of
the API clearly impact what bots can do.  GUI design choices also
could influence what types of bots people decide to implement.  Every
feature we observe in the Cryptsy interfaces plausibly has either a
neutral or positive effect on peer influence, and many of these
features are shared by other exchanges.  For both human and bot
traders, the features of the market mechanism and the ready
availability of recent trading activity, common to all the exchanges
we compared, is what enables peer influence in the first place.  For
human traders, the prominent display of trends in price history in the
ticker chart, again a feature common to all exchanges we compare,
plausibly encourages peer influence.  More uniquely, Cryptsy's
side-panel display of upward and downward price movements could have
led to a saliency bias \cite{hodas2012visibility}.  The chat
functionality on Cryptsy also likely promotes peer influence, although
of a different sort than the type we studied.  An important area of
future work is to more rigorously explore the causal impact of these
design features through laboratory experiments, field experiments, or
quasi-experimental observational analysis.


\subsection{Generalizability}

With the considerations of potential causes in mind, we expect our
results to likely generalize at a minimum to other cryptocurrency
exchanges, and possibly also at least to other online trading
platforms (ZuluTrade, eToro, etc.) and other small markets (e.g.,
penny stocks or pink sheets stocks).  The fact that Cryptsy
highlighted price change direction in the side-panel of its GUI is
relatively unique, but other exchanges have their own unique features
that could also promote peer influence, such as the audible beeps that
occur with all trades on Bitstamp.  Generalization to larger and more
professional financial markets is an important topic.  Similar effects
may be observable in higher-volume markets, but perhaps may require
larger or more sustained interventions to be detectable.  Obtaining a
more rigorous understanding of the interface effects in these larger
markets would be especially interesting.

\subsection{Design Implications}

In our discussion we have directed attention towards interface
features in particular as potential moderating contextual factors.  We
highlight the point that many of Cryptsy's design choices could have
plausibly promoted peer influence, and that Cryptsy could have
potentially made alternative choices that might inhibit it.  There is
an interesting moral hazard implicated in these
considerations. Cryptsy and other cryptocurrency exchanges make more
money when more people use the platform, so they are incentivized to
optimize their sites to stimulate trading, including via peer
influence.  At the same time, there are multiple factors that go into
creating a stable marketplace \cite{lampinen2017market}.  Maintaining
an awareness of incentives and a focus on design goals could help to
balance the positive and negative systemic effects of design choices.

\section{Related Work}




There are several bodies of work within computational social science,
economics and finance, and human-computer interaction related to the
present study.  We briefly review work on online field experiments in
peer influence, peer influence in financial markets, online collective
behavior, and the design of digital institutions.


\subsection{Online Field Experiments in Peer Influence}

A growing area across communities studying computational social
science is digital and online experimentation
\cite{bakshy2014designing,salganik2017bit}.  Our work was directly
inspired by earlier online experiments studying peer influence in a
variety of different types of online social systems
\cite{hanson1996hits,WattsMusic,bond201261,muchnik2013social,van2014field}. These
experiments have provided compelling evidence for the ubiquity of peer
influence across an array of domains, and in some cases the importance
of these effects on collective outcomes.  Ours is the first study to
apply the large-scale online field experimentation techniques
innovated in these early works to online financial markets.

\subsection{Peer Influence in Financial Markets}

There have been a number of observational studies, laboratory
experiments, and small-scale field experiments in finance and
economics studying peer influence in financial markets.  For instance,
the widely recognized empirical phenomenon of momentum in price
dynamics is related to our work.  A number of researchers have
identified evidence for momentum through observational analysis of
real financial markets
\cite{shiller1981stock,lo2002non,malkiel2003efficient} and through the
implementation of momentum-based trading strategies
\cite{jegadeesh1993returns,rouwenhorst1998international,jegadeesh2001profitability}.
In another line of work, models of ``herding'' formalize the behavior
of traders copying decisions to invest
\cite{banerjee1992simple,bikhchandani_theory_1992,avery1998multidimensional,chamley_rational_2004}.
These models have been tested directly in stylized laboratory
experiments
\cite{anderson1997information,cipriani2009herd,cipriani2005herd,drehmann2005herding}.
Herding has been argued to occur in real markets through observational
analyses of coarse-grained market data and individual institutional
investor data \cite{bikhchandani2000herd,spyrou2013herding}, but some
analyses with market data have yielded negative results
\cite{spyrou2013herding}.  The empirical work in finance and economics
most relevant to our own comes from the literature on attempting to
manipulate asset prices in laboratory asset markets
\cite{hanson2006information,veiga2009price,veiga2010information,buckley2015effect}
and certain types of real markets
\cite{CamererHorses,rhode2006manipulating}.  The general form of these
existing studies has been to execute large trades in the markets being
studied and observe the effects on market prices over a short time
period.  We focus on the effects of small individual trades rather
than attempts at market manipulation through abnormally large actions.

\subsection{Cryptocurrency Market Dynamics}

A growing body of work studies the dynamics of cryptocurrency markets
specifically.  Observational data analyses, with sometimes conflicting
results, have been used to examine competition between
cryptocurrencies for market volume \cite{gandal2014competition}; to
compute the fundamental value of cryptocurrencies
\cite{hayes2016cryptocurrency}; to investigate the efficiency of
bitcoin markets
\cite{urquhart2016inefficiency,nadarajah2017inefficiency}; to confirm
the presence of speculative behavior in bitcoin markets
\cite{macdonell2014popping,cheah2015speculative}; to provide evidence
for non-fundamentals-driven trading behavior in altcoin markets
\cite{elbahrawy2017evolutionary}; and to document how factors such as
fundamental value \cite{kristoufek2015main,bouoiyour2016drives} or
online search and discussion activity
\cite{kristoufek2013bitcoin,garcia2015social} are related to
cryptocurrency prices.  A recent study closely related in spirit to
our own provided observational evidence of market manipulation in
USD-BTC markets \cite{gandal2017price}.

\subsection{Studying Online Collective Behavior}


Our work also builds on a growing area within the human-computer
interaction (HCI) and computer-supported cooperative work (CSCW)
communities involving the study of human collective behavior in online platforms.
Some of this work has focused on financial markets
\cite{gilbert2010widespread,zhang2010trading}.  Peer influence is
related to the study of popularity dynamics in follower behavior
(e.g., \cite{hutto2013longitudinal}) and online voting behavior (e.g.,
\cite{lerman2010using,sipos2014review,leavitt2014upvoting}).  Others
have studied how design can help ameliorate the effects of peer
influence in online social recommendation systems (e.g.,
\cite{krishnan2014methodology,abeliuk2017taming}).  Our work also
employs the experimental technique of using bots in online field
experiments that has recently been developed in these communities
\cite{savage2016botivist,krafft2017bots}.

\subsection{Design of Digital Institutions}

A final related area of work is the design of digital institutions.
Many traditional institutions are shifting to having digital
components, and as this shift occurs, the study of institutions
becomes more relevant to researchers in human-computer interaction and
computer-supported cooperative work.  Researchers have studied a
variety of different types of institutions, for example: payment
systems \cite{Zhang2017ColdHE}; economics \cite{Bakshy2010TheSD} and
marriage \cite{Freeman2015SimulatingMG} in online worlds;
organizational behavior \cite{Grudin1988WhyCA}; activism
\cite{savage2016botivist}; knowledge markets \cite{Shen2012BarterMD};
labor markets
\cite{Shaw2011DesigningIF,Antin2012SocialDB,kittur2013future,Glss2016DesigningFL},
large-scale collaboration \cite{Benkler2012ThePA,Malone2017PuttingTP};
money (including cryptocurrencies)
\cite{Carroll2015CreatingVT,Sas2017DesignFT}; personal finance
\cite{gunaratne2015informing,gunaratne2017empowering}; supply chains
\cite{Pschetz2017}; entrepreneurship \cite{Jabbar2017GrowingTB};
hospitality \cite{Lampinen2016HostingVA}; and environmental
sustainability \cite{dourish2010hci}.  Broadly, these works bring a
design lens to bear on patterns of repeated digitally mediated
large-scale social interaction, and particularly a lens for how the
structure of technological artifacts affects our interaction patterns.
Others in the community have considered ethical
\cite{Irani2010PostcolonialCA,Vines2013ConfiguringPO,Brown2016FivePF,Dell2016TheIA,Harmon2017TheDF}
and conceptual \cite{malone1990coordination,Harvey2014HCIAA}
frameworks applicable to this type of design.  Specific interest in
HCI-oriented financial market design has existed since the early days
of online markets.  An early study investigated how market
interfaces can impact market liquidity compared to physical
``trading pits'' \cite{parikh1995electronic}.  Recent work has
emphasized a view of markets as technologically-mediated human
systems, and therefore a potentially fruitful target of HCI design and
critique \cite{lampinen2017market}.



\section{Conclusion}

Institutional design is a major area of study in the social sciences.
The {HCI} community has an opportunity to contribute to this
conversation as many farflung institutions---from banks to marriage in
Second Life---migrate to digital spaces.  The methodologies for
studying digital systems; the awareness of interface effects; and the
keen eyes for bias, ethics, and inclusion in the HCI community could
add unique perspectives to the design of digital institutions.
Markets are an example of an enormously important institution that is
becoming increasingly digitized, and market irrationality may be a
problem in markets that design-thinking could help address.  Our
specific application to cryptocurrencies is timely and urgent as new
platforms are growing and potentially encouraging users to adopt risky
trading strategies.

Our work provides an example of how peers in an online system can
audit the system dynamics through experimentation with typical
behavior.  Bots that randomly execute actions of normal users could
provide a way to understand peer influence and other phenomena in a
variety of online systems.  These bots allow us to understand the
causal impact of individual actions that can be taken by users in
these systems.  In studying the dynamics of cryptocurrency markets
with this technique, we show that even trades worth just fractions of
a penny can influence the nature of other much larger trades in the
cryptocurrency markets we study.  We observe an approximately two
percentage point increase in buying activity after our buy
interventions, and a cumulative monetary effect of 500 times the size
of our interventions.  While an increase of two percentage points
might seem small for an individual action, in a large marketplace this
amount is non-trivial.  For example, at the time of writing Apple
stock on the NASDAQ exchange had an average daily trading volume of 30
million USD, 2\% of which would amount to hundreds of thousands of
dollars over the course of a day on that stock alone.  Designers of
online markets should be aware of how minor changes in their systems
that affect individual and collective behavior could have major social
and economic impact.

\section{Appendix}

\subsection{Supplementary Analysis}


\subsubsection{Randomization Validity}

We assess the validity of our randomization procedure to check that
our treatment groups and control groups are not systematically
different.  We observe no detectable systematic differences between
treatment and control groups on the dependent variables we measure
before our interventions, but we do see significantly fewer treatment
observations than expected by chance (binomial test, $n = 2 \cdot
(52056), p < 1e-5$).  We attribute this difference to a failure to
treat, likely caused by a bug in Cryptsy that occasionally prevented
us from executing trades.  However, we find that our results are
robust to simulating these failures to treat in the control condition.
Plots of the numbers of control and treatment observations reveal
that no single coin is much more unbalanced than others in terms of
the number of control versus treatment
observations on that coin.

\subsubsection{Selection Bias}

There are several potential sources of bias in our experimental design
that could have influenced our estimated effect sizes.  We only
conducted interventions when the trade history and order book were
accessible via the Cryptsy API, when we had sufficient funds to buy
and sell at least 5e-6 of the coins they were monitoring, and when at
least one trade on the coin had occurred in the last hour.  The fact
that we condition on having access to the Cryptsy API, having enough
coins to trade, and having observed a trade at least an hour before
our interventions means that we are always conditioning on a
particular, albeit likely fairly common, market context in our
experiments.  Another source of bias in two of our dependent variables
is missing trials due to having no trades observed in our monitor
windows.  Both the probability that the last observed trade is a buy
and the percentage of buy-side volume are undefined when no trades are
observed.  We achieved only approximately 50\% overall probability of
observing any trades within 15 minutes after our interventions, and
this probability varies widely across coins. The combined effects of
these two sources of selection bias can be observed by looking at the
number of observations we have per coin.  Since higher volume coins
are likely to meet both our condition for intervention and our
condition for measurement, any bias on the marginal effects we examine
will be towards the effect on higher volume coins, which are of more
general interest anyway.  Regardless, the fixed effects regressions we
performed helps to control for the selection bias due to coin-level
effects.  Since our treatments affect the observability of our outcome
variables, we might also be concerned that observability alone is
driving our effects.  However, we see that there is a significant
difference in effect between buy and sell treatments, which do not
differ strongly in terms of observability.  In this analysis, the treatment type does
not significantly interact with pre-treatment market state variables
when predicting observability and controlling for multiple comparison
in a linear regression.



\subsubsection{Heterogeneous Effects}

We observe substantial variation across coins in our dependent
variables and in the effect sizes of our interventions.  However, we
did not identify any descriptive statistics of the coins that were
reliably predictive of effect size (with effect size measured by the
difference between the treatment mean and the control mean).  A
multivariate linear regression including median values of coin
attributes (price, volume, spread, and best open sell/buy order size)
yielded no significant relationships with effect size and low
R-squared values of approximately 0.01.  Since much of the variability
occurs on coins with fewer observations, the major variations across
coins are therefore likely largely due to noise.  We needed a large
sample size in order to be able to detect aggregate peer influence
effects in these markets at all.  We appear to have too small of a
sample size per individual altcoin, and too few coins, to examine
heterogeneous treatment effects.

\subsection{Code, Data, Preregistration, and Institutional Review}

Our code and data, including a list of cryptocurrencies we traded, is
available online:
\url{https://github.com/pkrafft/An-Experimental-Study-of-Cryptocurrency-Market-Dynamics}.
Code for our experiment specifying  the
details of our experimental design was
preregistered.\footnote{\url{https://osf.io/djezp/}} Our final
statistical analysis differed from our preregistered one.
Before conducting
non-preregistered data analysis, we split our entire dataset of trials
uniformly at random into two halves, one for exploratory statistical
analysis and one for confirmatory statistical analysis.  The results
we show are from our confirmatory validation set.  This confirmatory
set was held-out from analysis as much possible.  Our experiment was
approved by the human subjects review boards of MIT and ANU (MIT
Protocol \#1409006623, ANU Protocol: 2015/652).  Because our
experiments were conducted in the field and consisted of no more than
regular trading activity in the markets we used, our study posed
minimal risk to our participants and therefore was granted a waiver of
informed consent.

\section{Acknowledgments}

This material is based upon work supported by the NSF Graduate
Research Fellowship under Grant No. 1122374, the MIT Media Lab Members
Consortium, the Australian National University, Harper Reed, and Erik
Garrison.  Any opinion, findings, and conclusions or recommendations
expressed in this material are those of the authors(s) and do not
necessarily reflect the views of our sponsors.  Statistical support
was provided by data science specialists Simo Goshev and Steven
Worthington at the Institute for Quantitative Social Science, Harvard
University.  Special thanks to Iyad Rahwan for suggesting the moral
implications of our results, and to the participants of the Second EC
Workshop on Crowdsourcing and Behavioral Experiments for their
feedback on an early version of this work.


\balance{}

\bibliographystyle{SIGCHI-Reference-Format}

\bibliography{cryptsy}

\end{document}